\documentclass[seceq]{ptptex}

\usepackage{graphicx}
\usepackage{wrapft}
\usepackage{epsfig}

\newcommand{\nudyn}{$\nu_{(+-,dyn.)}$}

\newcommand{\deltaeta}{$\Delta \eta$}
\newcommand{\nchnudyn}{$\nu_{(+-,dyn.)} \times N_{ch}$~}

\title{
First event--by--event fluctuation studies in Pb-Pb collisions at LHC energy by the ALICE experiment%
}

\author{
Satyajit \textsc{Jena} for the ALICE collaboration%
}

\inst{
Indian Institute of Technology Bombay, Mumbai-76, India
}

\abst{The presence of the phase transition can manifest itself by the 
characteristic behavior of several observables which may vary 
dramatically from one event to the other. Thus, the study of 
various conserved quantities on an event-by-event basis  
offers the possibility to study the phase transition and the nature 
of high density matter. The ALICE experiment is well suited for  
precise event-by-event measurements of various quantities. 
In this article, the event-by-event fluctuations of mean transverse 
momentum and net-charge distributions as measured by the ALICE experiment 
are presented.
}

\begin{document}
\maketitle


\section{Introduction}
Heavy-ion collisions at ultra-relativistic energies
can produce a new state of matter, which is characterized by high
temperature and energy density, where the degrees of freedom are given no more 
by the hadrons but by their constituents, the quarks and the gluons.
The ALICE experiment~\cite{Aamodt08}, located at 
the CERN LHC, is a multi-purpose experiments with highly sensitive detectors 
around the interaction point. The central detectors cover the 
pseudorapidity region $|\eta| < 0.9$, 
with good momentum measurement as well as good particle identification capabilities.
This gives us an excellent opportunity to study the fluctuations and 
correlations of physical observables on an event-by-event basis.
In this analysis, the Time Projection Chamber (TPC)~\cite{tpc} is used for selecting 
tracks, the Inner Tracking System (ITS) is used for vertexing and triggering and
VZERO scintillator hodoscope is used for estimating centrality~\cite{toia} as 
well as triggering. The present analysis is performed by taking the z-vertex 
within $\pm$10 cm of the detector's center, $|\eta| < 0.8$ and the charged 
particle transverse momentum, $p_{\rm T}$, from 0.15 GeV/c to 2 GeV/c.

\section{Mean transverse momentum fluctuations}
Event-by-event fluctuations of mean transverse momentum, i.e. 
$\langle p_{\rm T} \rangle$, contain information on the 
dynamics and correlations in heavy-ion collisions.
The fluctuations in $\langle p_{\rm T} \rangle$ may 
be related to the critical behavior of the system in the 
vicinity of a phase boundary \cite{stephanov1,stephanov2} or the occurrence of 
thermalization and collectivity \cite{gavin}. In general, 
there may be a variation of fluctuations (increase or decrease) in heavy-ion collisions
with respect to the pp collisions which serves as the baseline measurement. 

The two-particle correlator~\cite{voloshin} 
$C_m = \langle \Delta p_{{\rm T},i}, \Delta p_{{\rm T},j} \rangle$ 
is a measure of the dynamical component of the variance of $\langle p_{\rm T} \rangle$ 
and is defined by
\begin{equation}
  C_m = \frac{1}{\sum_{k=1}^{n_{\rm ev}}{N_{k}^{\rm pairs}}} \cdot 
  \sum_{k=1}^{n_{\rm ev}} \sum_{i=1}^{N_{k}} \sum_{j=i+1}^{N_{k}} (p_{{\rm T},i} - 
  \langle p_{\rm T} \rangle_m ) \cdot (p_{{\rm T},j} - \langle p_{\rm T} \rangle_m ),
  \label{eq:correlator}
\end{equation}
where $n_{\rm ev}$ is the number of events in a given multiplicity 
class $m$, $N_{k}^{\rm pairs}$ is the number of pairs constructed out of $N_k$ number
of particles in an event and equal to $0.5 \cdot N_{k} \cdot (N_{k}-1)$ and
$\langle p_{\rm T} \rangle_m$ is the average $p_{\rm T}$ of all tracks of all events 
in class $m$. A null value of $C_m$ is obtained in the presence of only 
statistical fluctuations. 
\begin{figure}
 \vskip -0.1cm
\centering
\includegraphics[width=5.8 cm]{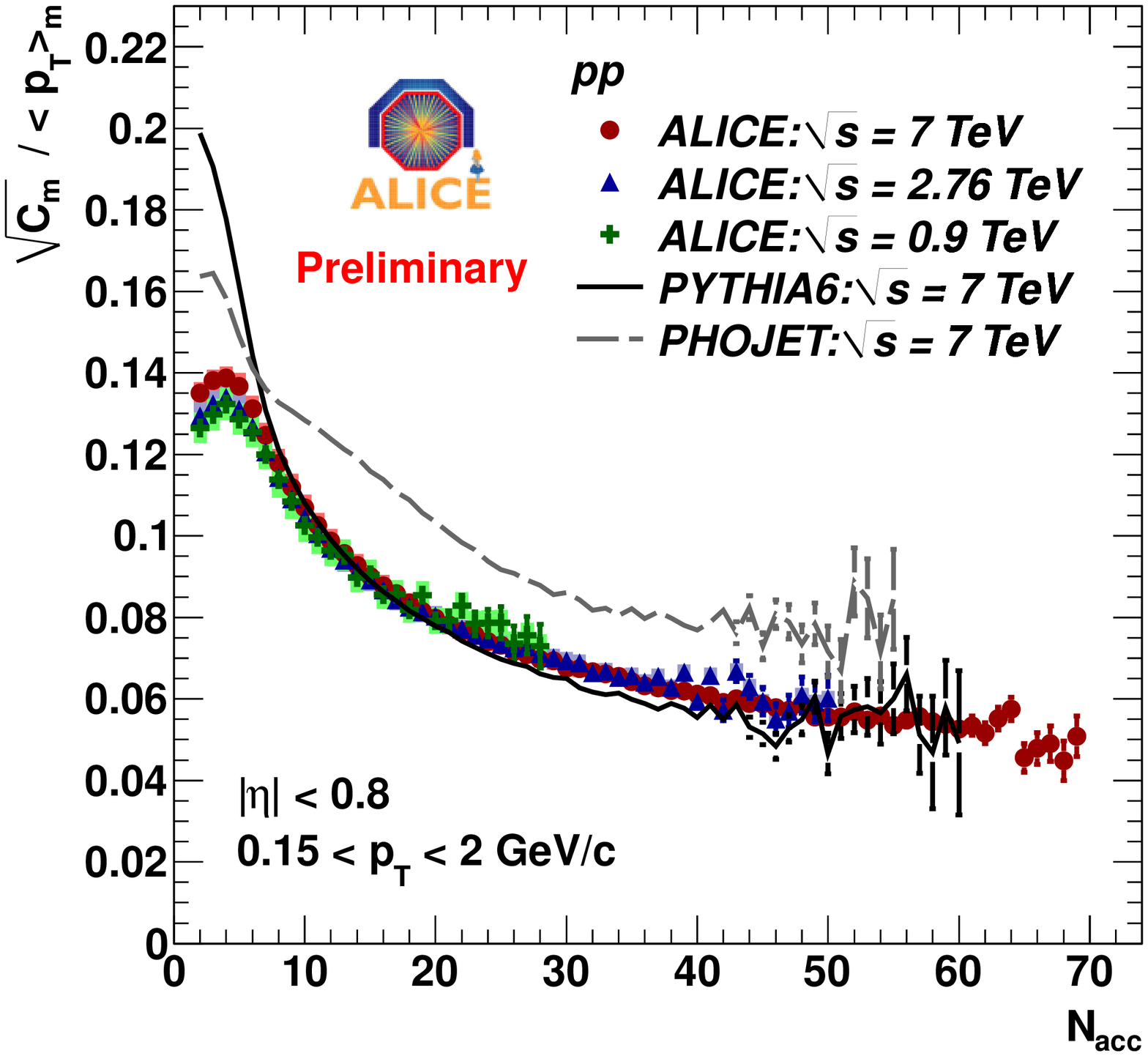}
\includegraphics[width=5.8 cm]{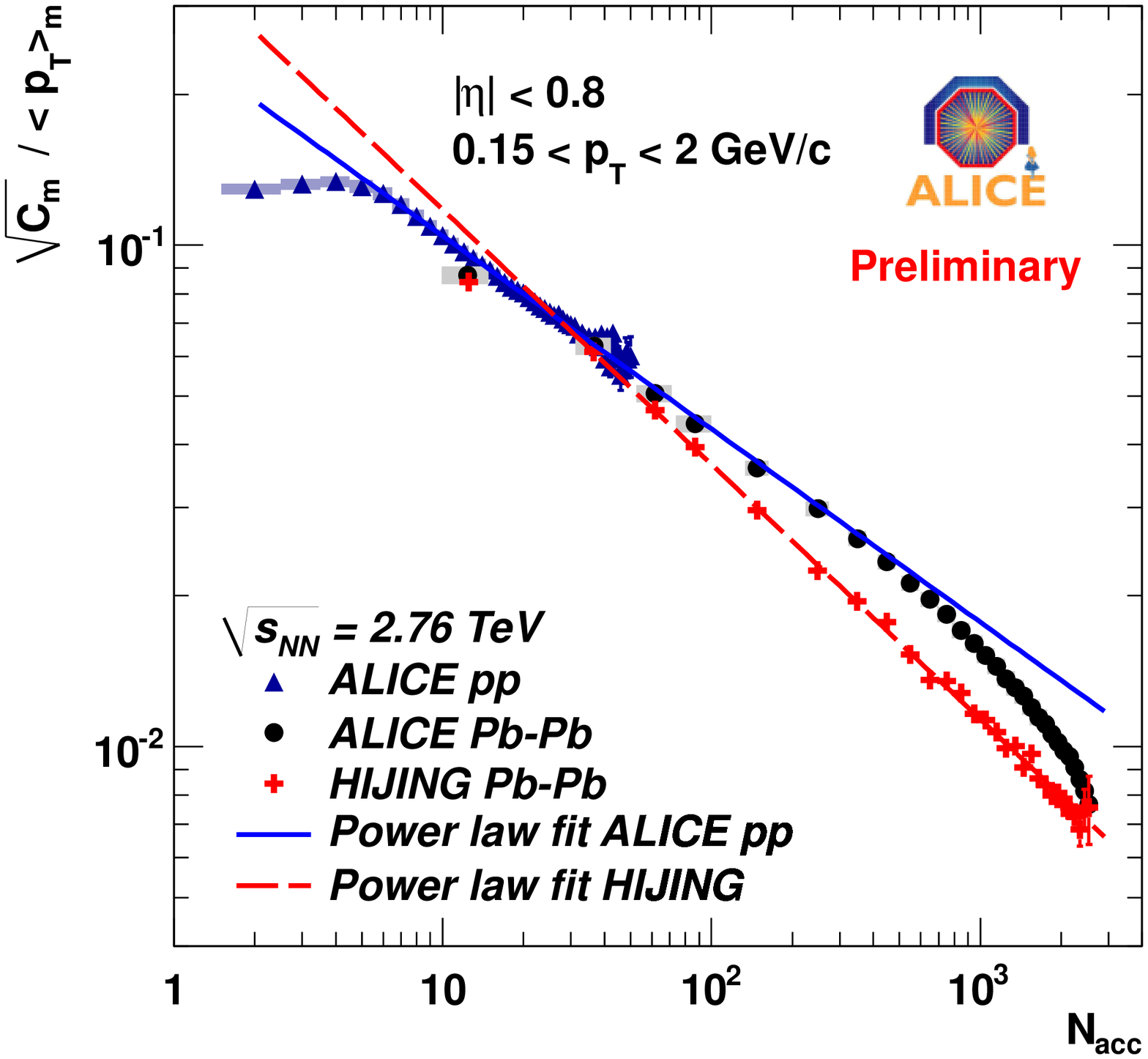}
\caption{Relative fluctuations $\sqrt{C_m} / \langle p_{\rm T} \rangle_m$ 
as a function of accepted multiplicity $N_{\rm acc}$, for
pp collisions compared with MC simulation(left) and pp with Pb-Pb collisions(right). 
See text for details.}
\label{fig:PtFluc_figure1}
\vskip -0.2cm
\end{figure}

In figure~\ref{fig:PtFluc_figure1}, the relative fluctuation, i.e.
$\sqrt{C_m} / \langle p_{\rm T} \rangle_m$, is plotted as a function of
accepted number of tracks ($N_{\rm acc}$).
In the left panel of figure~\ref{fig:PtFluc_figure1}, the
 results for pp collisions
at $\sqrt{s}$ = 0.9, 2.76, and 7 TeV are presented.
The p-p result at $\sqrt{s}$ = 7 TeV is compared
with the PYTHIA6~\cite{pythia} and PHOJET~\cite{phojet} models at the same energy. 
It is observed that PYTHIA6 (Perugia0 tune) describes the data reasonably well 
above $N_{\rm acc} \geq$ 7, whereas PHOJET 
overpredicts the data. 
\begin{wrapfigure}{r}{8.8cm}
   \vskip -0.2cm
  \includegraphics[width=0.64\textwidth]{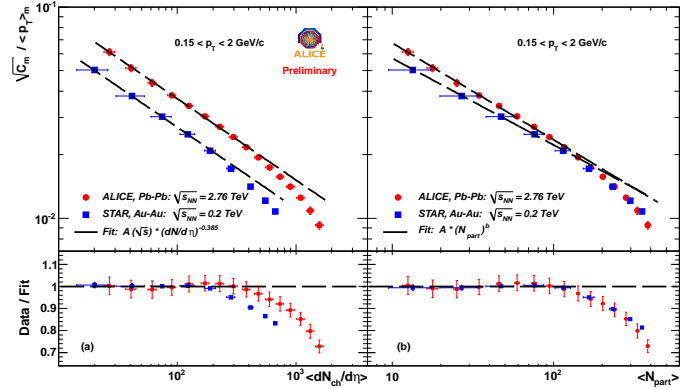}
  \caption{Relative fluctuations in Pb--Pb collisions measured by ALICE 
    and Au-Au collisions measured by STAR as a function of $\frac{dN_{ch}}{d\eta}$
    (a) and $N_{\rm part}$ (b).}
  \label{fig:PtFluc_figure3}
  \end{wrapfigure}
It is also clear from the figure that there exist a 
significant dynamical fluctuations in the data. 
In the right panel of figure~\ref{fig:PtFluc_figure1}, Pb--Pb data 
at $\sqrt{s}_{NN}$ = 2.76 TeV along with
the pp data are presented. The non-statistical
fluctuations decrease with the increase of multiplicity.
The pp data are fitted with a power law 
$\sqrt{C_m} / \langle p_{\rm T} \rangle_m (N_{\rm acc}) = A \cdot N_{\rm acc}^{b}$, 
with $b = -0.385\pm0.003$. 
The Pb--Pb data agree well with this pp baseline 
up to $N_{\rm acc} \approx 600$, justifying the use of a common 
parameterization. 
Central Pb--Pb collisions show a significant additional reduction of 
the fluctuations. 
It is found that the 
HIJING~\cite{hijing} data do not describe the shape of the Pb--Pb.

In figure~\ref{fig:PtFluc_figure3}(a), the Pb--Pb data of ALICE 
and Au--Au data of STAR~\cite{star} are fitted
with a power law with parameter $b$ = -0.385, which is obtained from the ALICE
pp data at $\sqrt{s}$ = 2.76 TeV. Both agree with the same 
parametrization in peripheral events, suggesting the validity of the pp baseline
for RHIC A--A data. 
The deviation increases with multiplicity and is more in Pb--Pb data 
compared to Au--Au data. In figure~\ref{fig:PtFluc_figure3}(b), the relative 
fluctuations of STAR and ALICE data are presented
as a function of $N_{part}$. Central collisions in both datasets are not 
described by the power law (with a common exponent for ALICE and STAR) 
obtained from fits to the peripheral. 
\section{Fluctuations of the net charge}
The fluctuations of net--charge depend on the squares of the charge states 
present in the system. The QGP phase, having  quarks as the charge 
carriers, should result into a significantly different magnitude of 
fluctuation compared to a hadron gas (HG). Net--charge fluctuations may be 
expressed by the quantity $D$, defined as~\cite{JeonKoch00}:
\vskip -0.8cm
\begin{eqnarray}
D = 4 \frac{\langle \delta Q^2\rangle}{N_{\rm ch}}  \approx \nu_{(+-,dyn)}  \times \langle N_{\rm ch} \rangle + 4
\end{eqnarray}
\vskip -0.3cm
\noindent where $\langle \delta Q^2 \rangle$ is the variance of the 
net--charge Q with $Q = N_+ - N_-$ and $N_{\rm ch} = N_+ + N_-$. Here $N_+$ 
and $N_-$ are the numbers of positive and negative particles. 
And the $\nu_{(+-,dyn)}$, independent of detector acceptance and efficiencies 
is defined by
\vskip -0.5cm
\begin{eqnarray}
\nu_{(+-,dyn.)} = 
 \frac{\langle N_+(N_+-1) \rangle}{\langle N_+ \rangle ^2} +
 \frac{\langle N_-(N_--1) \rangle}{\langle N_- \rangle ^2} 
- 2\frac{\langle N_-N_+ \rangle}{\langle N_- \rangle \langle N_+ \rangle},
\end{eqnarray}
\begin{wrapfigure}{r}{6.6cm}
\vskip -0.8cm
\begin{center}
\includegraphics[width=\linewidth]{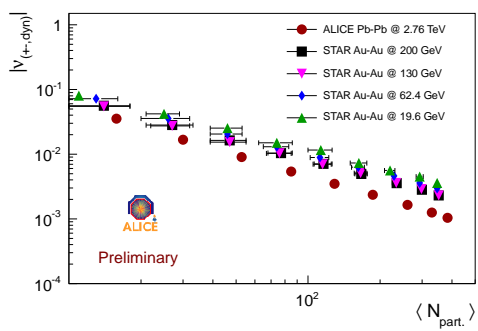}
\caption{The absolute value of $\nu_{+-,dyn}$,  as a function of 
 the collision centrality compared with measurements for lower energies.}
\label{fig2Log}
\end{center}
\end{wrapfigure}

Figure~\ref{fig2Log} represents the absolute value of the dynamical 
net--charge fluctuations, \nudyn, as a function of number of participants 
in  \mbox{Pb--Pb} collisions at $\sqrt{s_{NN}}=2.76$~TeV and Au+Au 
collisions at $\sqrt{s_{NN}}=$~ 19.6, 62.4, 130, and 200 GeV at STAR. It is observed
that the dynamical net--charge fluctuations exhibit a monotonic 
dependence on the number of participating nucleons. 

\begin{figure}[htb]
\centering
    \includegraphics[width=6.6cm, height=4.5cm]{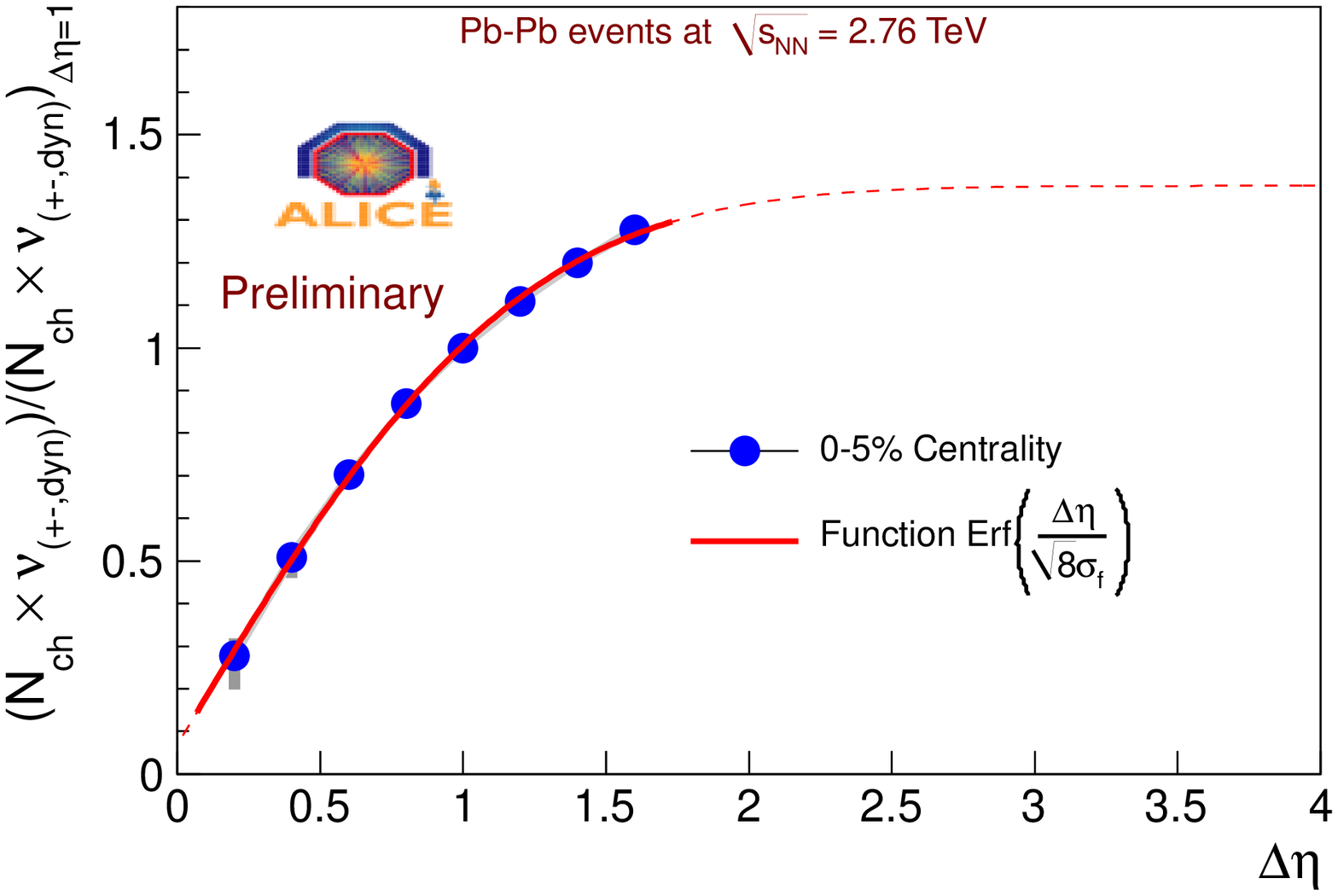}
    \includegraphics[width=6.6cm, height=4.5cm]{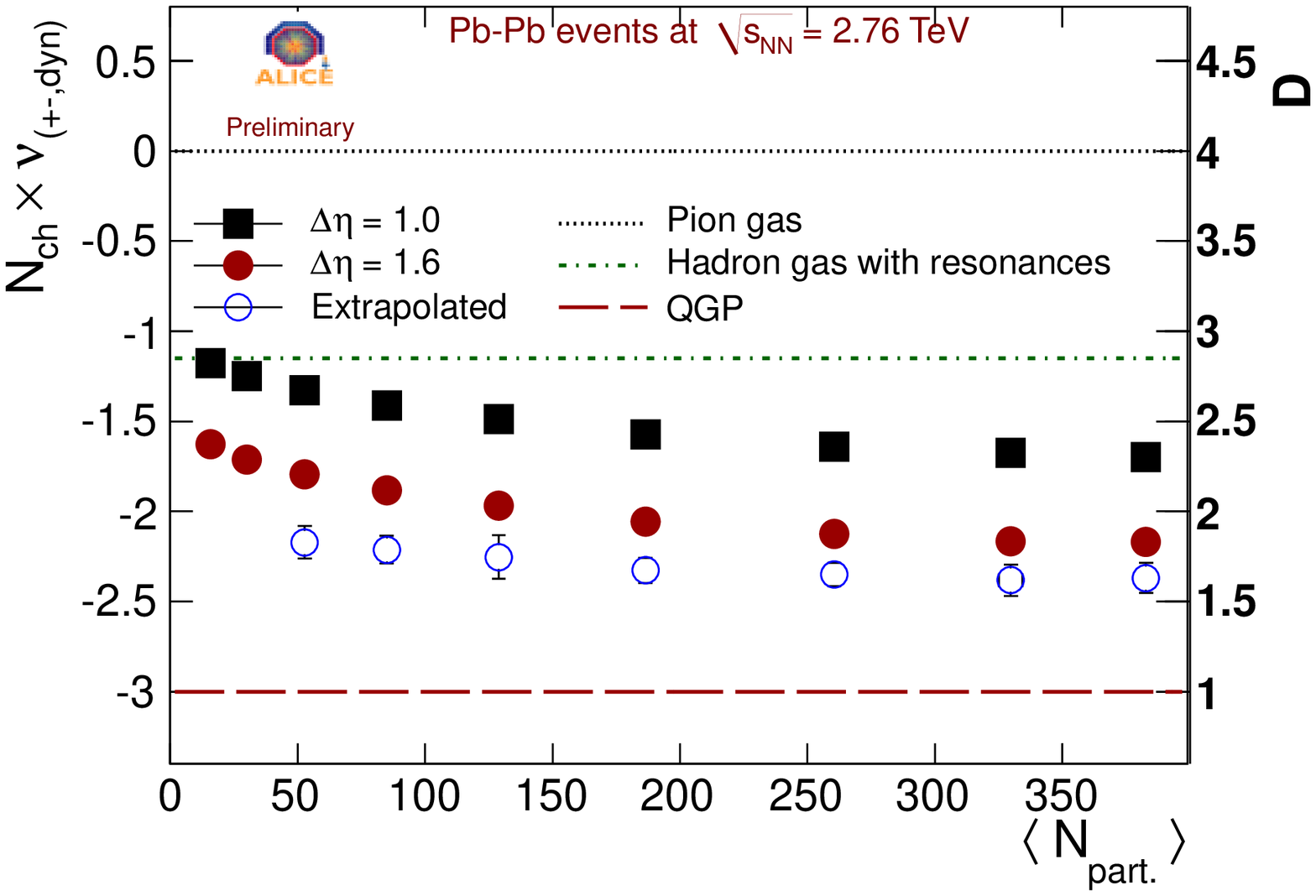}
    \caption{(left) ${\rm N}_{ch}\times\nu_{(+-,dyn)}$, normalized 
      to the values for $\Delta\eta=1$, as a function of \deltaeta. 
      (right) ${\rm N}_{ch}\times\nu_{+-,dyn}$,  as a function of 
      the collision centrality for \deltaeta~ values of 1.0, 1.6 and
      extrapolated values.
\label{fig5}
} 
\end{figure}

In left panel of figure~\ref{fig5}, the nature of the variation of 
\nchnudyn with \deltaeta~ is studied by plotting its ratio with respect to 
the value at $\Delta\eta = 1$. It is observed that the relative value 
of \nchnudyn~grows smoothly with 
increasing \deltaeta~ window. This behavior has been predicted  
earlier \cite{Shuryak01,AbdelAziz05} and was attributed to the 
dissipation of the signal arising from hadronic diffusion during the
evolution from quark gluon plasma stage to
the freeze-out stage. The data points are fitted with an error function 
of the form erf($\Delta\eta/\sqrt{8}\sigma_f$), representing the diffusion 
process. The diffusion coefficient, $\sigma_f$, obtained from the fitting 
is equal to $ 0.467 \pm 0.021$ at 0-5\% centrality.

In the right panel of figure~\ref{fig5}, the net--charge 
fluctuations, expressed in terms of \nchnudyn and $D$ (left-- and 
right--axis, respectively) as a 
function of the $N_{part}$ are shown for three different 
\deltaeta~windows, i.e. $\Delta \eta = 1$, $\Delta \eta = 1.6$ and the 
extrapolated asymptotic values at $\Delta \eta = 3$. By confronting the measured 
value with the theoretically predicted fluctuations \cite{JeonKoch00, Shuryak01}, 
it is observed that the results are within the limits of the QGP and the 
HG scenarios. Since the fluctuations should normally grow in the 
process of the evolution of the system till freeze--out, the value 
obtained by the experiment might have its origin from a QGP phase.

In summary, the fluctuations of mean transverse momentum get diluted 
for higher multiplicities. Peripheral A-A data both from ALICE and 
STAR agree well with the ALICE pp baseline. For central A-A collisions, 
an additional reduction of the fluctuations is observed. The net-–charge 
fluctuations 
are observed to be dominated by the correlations of oppositely charged 
particles. The energy dependence of the dynamical fluctuations shows 
a decrease in fluctuation going from RHIC to LHC. The measured dynamical 
fluctuation is closer to the theoritically predicted value of 
Quark-Gluon Plasma.

%

\end{document}